\documentclass[11pt]{article}
\usepackage{amsmath}
\usepackage{amsfonts}
\usepackage{amssymb}
\usepackage{graphicx}
\usepackage[plainpages=false,pdfpagelabels,pagebackref=false,naturalnames=true,hyperindex=true,pdftitle={Is the world digital or analog},pdfauthor={Hector Zenil}]{hyperref}

\newenvironment{myenumerate}{
\begin{enumerate}
  \setlength{\itemsep}{1pt}
  \setlength{\parskip}{0pt}
  \setlength{\parsep}{0pt}}{\end{enumerate}
}

\newenvironment{myitemize}{
\begin{itemize}
  \setlength{\itemsep}{1pt}
  \setlength{\parskip}{0pt}
  \setlength{\parsep}{0pt}}{\end{itemize}
}

\begin{document}

\date{}

\title{Image Characterization and Classification\\by \emph{Physical} Complexity}
\author{\medskip Hector~Zenil$^\dagger$,~Jean-Paul Delahaye$^\dagger$~and~C\'{e}dric~Gaucherel$^\ddagger$\\$^\dagger$LIFL - UMR CNRS 8022, Lille, France\\$^\ddagger$INRA, UMR AMAP, Montpellier, F-34000 France}

\maketitle

\begin{abstract}

We present a method for estimating the complexity of an image based on Bennett's concept of logical depth. Bennett identified logical depth as the appropriate measure of organized complexity, and  hence  as being better suited  to  the evaluation of the complexity of objects in the physical world. Its use results in a different, and in some sense a finer characterization than is obtained through the application of the concept of Kolmogorov complexity alone. We use this measure to classify images by their information content. The method provides a means for classifying and evaluating the complexity of objects by way of their visual representations. To the authors' knowledge, the method and application inspired by the concept of logical depth presented herein are  being  proposed and implemented for the first time.\\

\end{abstract}


\section{Introduction}
We present a method for assessing and quantifying the information content of an image based on the theory of algorithmic information, specifically on Bennett's concept of logical depth \cite{bennett}.  It serves as a means for evaluating and classifying images by their organized complexity. Images have a number of features containing information in the form of pixels. As a representation of an object, an image constitutes a description of said object, capturing some of its characteristics.

 Algorithmic information theory \cite{kolmogorov,chaitin} formalizes the concepts of simplicity and randomness by means of information. Many applications of the theory of algorithmic information have been developed to date, for example  \cite{chen,goel,li2,li3,zenil}. For a detailed survey see  \cite{li,cilibrasi2}. None, however, seems to have exploited the concept of logical depth, which may provide another useful complexity measure. Logical depth was originally identified with what is usually believed to be the right measure for evaluating the complexity of real-world objects such as living beings. Hence its alternative designation: physical complexity (used by Bennett himself). This is because the concept of logical depth takes into account the plausible history of an object. It combines the shortest possible description of the object with the time that it takes for this description to evolve to its current state. The addition of logical depth results in a reasonable characterization of the organizational (physical) complexity of an object, which is not to be had by the application of 

the concept of algorithmic complexity alone. The main hypothesis of this paper is that images can be used to determine the physical complexity of an object (or a part of it) \cite{gaucherel} at a specific scale and level, or if preferred, to determine the complexity of the image containing information about an object. And as such, they cannot all be presumed to have the same history, but rather to span a wide and measurable spectrum of evolutions leading to patterns with different complexities, ranging from the random-looking to the highly organized. To test the applicability of the concept of logical depth to a real-world problem, we first approximate the shortest description of an image by way of currently available lossless compression algorithms. Then the decompression times are estimated as an approximation to the logical depth of the image. This allows us to generate a relative measure and to produce a classification based on this measure.

 The paper is organized as follows: In section \ref{theo}  the theoretical background that will constitute the formal basis for the proposed method is introduced. In Section \ref{method} we describe the method and the battery of tests used to evaluate it. Finally, in section \ref{results} we present the results followed by the conclusions in \ref{conclusions}.

\section{Theoretical foundation}
\label{theo}

\subsection{Algorithmic complexity}

The complexity of a string of bits can be defined in terms of algorithmic complexity\footnote{Also known under the names program-size complexity and Kolmogorov-Chaitin complexity.}. That is, given a program producing a string s, a machine can run the program and make a copy of $s$ (in our case the image of the represented object). The information content of a bit string can be defined as the length (in bits) of the smallest program which produces the string $s$. 

In algorithmic information theory a string is algorithmically random if it is incompressible. The difference in length between a string and the shortest algorithm able to generate it is the string's degree of complexity. A string of low complexity is highly compressible, as the information that it contains can be encoded in an algorithm much shorter than the string itself, while a string of high complexity is hard to compress because, in a fixed language, its shortest possible description is itself.

The algorithmic complexity \cite{kolmogorov,chaitin,levin,solomonoff} $K_U(s)$ of a binary string $s$ with respect to a universal Turing machine $U$ is defined as the binary length of the shortest program $p$ of length $|p|$ that produces $s$ as output:

\begin{center}
$K_U(s) = \min \{|p|, U(p)=s\}$
\end{center}

$K$  however is not a computable function. In other words, there is no program which takes a string $s$ as input and produces the integer $K(s)$ as output. Since $K(s)$ is the length of the shortest compressed form of $s$, that is, the best possible compression (up to an additive constant), one can approximate $K$ by compression means, using currently available lossless compression programs. The length of the binary compressed version of $s$ is an upper bound of its algorithmic complexity and therefore an approximation of $K(s)$. Clear introductions to the subject are available in  \cite{calude,li}.

\subsection{Bennett's logical depth}

A measure of the complexity of a string can be arrived at by combining the notions of algorithmic information content and time. According to the concept of logical depth \cite{bennett,bennett2}, the complexity of a string is best defined by the time that an unfolding process takes to reproduce the string from its shortest description. The longer it takes, the more complex. Hence complex objects are those which can be seen as ``containing internal evidence of a nontrivial causal history." 

Unlike algorithmic complexity, which assigns a high complexity to both random and highly organized objects, placing them at the same level, logical depth assigns a low complexity to both random and trivial objects. It is thus more in keeping with our intuitive sense of the complexity of physical objects, because trivial and random objects are intuitively easy to produce, have no long history, and unfold quickly. A persuasive case for the convenience of the concept of logical depth as a measure of organized complexity, over and against algorithmic complexity used by itself, is made in  \cite{delahaye}. Bennett's main motivation was actually to provide a reasonable means of measuring the physical complexity of real-world objects. 

Bennett provides a careful development \cite{bennett} of the notion of logical depth, taking into account near-shortest programs as well as the shortest one--hence the significance value--to arrive at a reasonably robust measure. For finite strings, one of Bennett's formal approaches to the logical depth of a string is defined as follows:

Let $s$ be a string and $d$ a significance parameter. A string's depth at significance $d$, is given by

\begin{center}
$D_d(s)$=$\min\{T(p) : (|p|-|p^\prime| < d) \wedge (U(p) = s)\}$
\end{center}

\noindent  the number of steps $T(p)$ in the computation $U(p)=s$, with $|p^\prime|$ the length of the shortest program for $s$, (therefore $K(s)$). In other words, $D_d(s)$ is the least time $T$ required to compute $s$ from a $d$-incompressible program $p$ on a Turing machine $U$. 

The simplicity of Bennett's first definition  in  \cite{bennett} makes it suitable for an initial investigation, serving as a practical approximation to this measure via the decompression times of compressed versions using lossless compression algorithms. The decompression time of the compressed version of a string is a lower bound of Bennett's logical depth. From now on what we have identified as the measure approximating the logical depth of a string $s$ will be denoted by $D(s)$, with no need of a significance parameter due to the fact that we are using Bennett's first, simpler definition.

 Algorithmic complexity and logical depth are intimately related. The latter depends on the former because one has to first find the shortest programs producing the string and then look for the shortest times (Looking for the shortest programs is equivalent to approximating the algorithmic complexity of a string.). While the compressed versions of a string are approximations of its algorithmic complexity, decompression times are approximations of the logical depth of the string. These approximations depend on the compression algorithm used. For algorithmic complexity the choice of universal Turing machine is bounded by an additive constant (invariance theorem) while for logical depth the choices involved are bounded by a multiplicative factor \cite{bennett}.

\section{Methodology}
\label{method}

An image can be coded as a string $s$ over a finite alphabet, say the binary alphabet--a  black and white image. We will denote by $K_c$ and $D_c$ the approximations obtained by means of a compression algorithm $c$. When the algorithmic complexity $K(s)$ is approximated by a lossless compression algorithm, this approximation corresponds to an upper-bound of $K(s)$ \cite{li}.

 There seems to have been no previous attempt to implement an application of ideas based on Bennett's logical depth to a real-world problem. In order to assess the feasibility of an application of the concept to the problem of image characterization and classification by complexity, we performed a series of experiments of gradually increasing  sophistication, starting from fully controlled experiments and proceeding to the use of the best known compression algorithms over a larger dataset.

 The battery of tests involved a series of images devised to tune and verify different aspects of the methodology, and a more realistic dataset, indicating whether the results were stable enough to yield the same values after each repetition of the experiment and whether they were consistent with the theory and consonant with an intuitive sense of a complex vs. a simple object. 

The first experiments consisted in controlling all the environmental parameters involved, from the data to the compression algorithm, in order to test our first attempts to calculate decompression times. A test to measure the correlation between image sizes (random vs. uniformly colored images) and decompression times was carried out. The results are in section \ref{size}. 

A second test in section \ref{cuadros} consisted of a series of images meant to evaluate the change and magnitude of the decompression times when manipulating the internal structure of an image. That is, it served to verify that the decompression times decreased as expected when the content of a random image was artificially manipulated to make it more simple. This consisted of a set of images in which uniform structures consisting of large single-bit strings of a fixed size were randomly inserted into images originally containing only pseudo-randomly generated pixels.

 The framework entailed the use of a toy compression program involving an algorithm 
grouping runs of the same bit replaced by a couple of values: the replaced bit followed by the number of times the bit was found. No further allowances were made, either for dealing with special cases or for detecting any other kinds of patterns. We wanted the algorithm and the data to be as simple as possible, to be fully controlled in every detail so we could better understand the relation between an increase in the size of a structure and  the decompression time. It consisted of the series of computer-generated random images shown in Figure 1.

\begin{figure}[h!]
  \centering
      \scalebox{.75}{\includegraphics{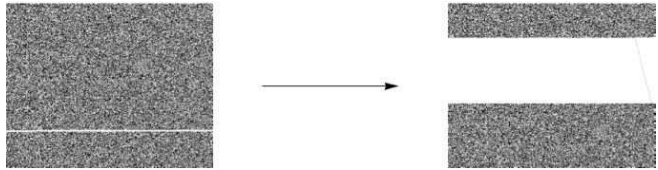}}
  \caption{Inserting increasingly larger (white) regular blocks into an image originally containing randomly distributed black and white pixels.}
\end{figure}

This was a way of injecting structure into the images in order to study the behavior of the compression algorithm and ensure control of the variables involved so as to better understand the next set of results. 

A procedure described in \ref{lineas} was devised to test the case complementary to the previous test-- to verify that the decompression times increased when the content of an image was artificially manipulated, transforming it from a simple state (an all-white image) to a more complex state by generating images with an increasing number of structures. A collection of one hundred images with an increasing number of straight lines randomly inserted was artificially generated (see Figure 2). The process led to interesting results described in \ref{lineas}, demonstrating the robustness of the method.

\begin{figure}[h!]
  \centering
      \scalebox{.9}{\includegraphics{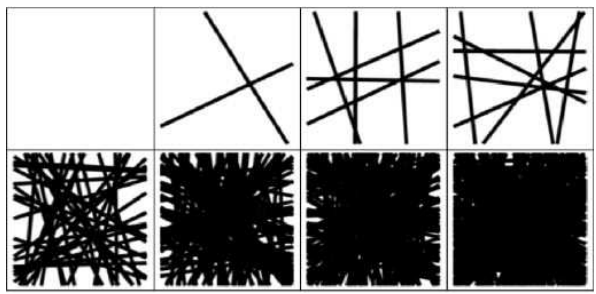}}
  \caption{The number of lines grows as $2n^2$ with $n$ the image number from $0$ to $99$. Introducing lines at this rate allows us to start with a uniformly white image and end up with the desired, nearly all-black uniformly colored image in 100 steps.}
\end{figure}

Three more tests to calibrate the measure based on logical depth are proposed in \ref{calibration}, with three different series of images. The first two series of images were computer generated. The third one was a set of pictures of a real-world object. The description of the 3 series follows:

\begin{myenumerate}
\item A series of images consisting of random points converted
 to black and white with a probability of a point's existing at any given location having a 
certain value depending on a threshold. This can be seen as variation in information content, since the ratio of black and white dots varies from high to low from image to image depending on the threshold value. 
\item Cellular automaton with elementary rule number 30 (in Wolfram's enumeration \cite{wolfram}), then another image with the same automaton superposed upon itself (rotated 90${}^{\circ}$), followed by the same automaton to which was applied a function introducing random bits to 50 percent of the image pixels. We expect a greater standard deviation for the series of cellular automata because they come in pairs (meaning each image comes with its inverse), and permutations should be common and ought to occur between these pairs because, intuitively, they should have the same complexity. This guess will be statistically confirmed in the results\ref{results} section.
 \item A wall and the same wall but viewed from a closer vantage point. 
\end{myenumerate}

The last test in \ref{main} is the main result of the paper, showing the stability of the method and the classification result. The experiment was performed on a dataset of randomly chosen pictures, 56 black-and-white pictures coming from different sources and representing objects of all kinds\footnote{The images are available online under the paper title at \url{http://www.mathrix.org/experimentalAIT/}}. Some were pictures of actual objects produced by nature and by humans, such as car plans, pictures of faces, handwriting, drawings, walls, insects, and so on. Others were computer-generated, such as straight lines, Peano curves, fractals, cellular automata, monochrome and pseudo-randomly generated images.

 The selection of images in section \ref{main} was made by hand, bearing in mind the goal of including a large variety of objects covering a range of seemingly different complexities. We chose images of faces, people, engines and electronic boards, images singled out for their high degree of  complexity, each being usually the result of a relatively long history, whether human artifacts or long-lived natural entities. The image bearing the name ``table" is a table of numerical values-- a computer spreadsheet-- likely characterized by a significant level of complexity. Images tagged ``people" are pictures of a group of people taken from a distance. The tag ``inv" following picture names indicates that they are the color inversions of other images in the same dataset that we expected would be close in complexity, and they are presented side-by-side with their non-inverted versions. ``Writing" refers to handwriting by a human being, which should also have a significant level of complexity, a level of complexity approaching that of other handwritten pieces such as the image tagged  ``formulae," which depicts formulae  rather than words. ``Watch" is the internal engine of a watch. ``Cpu A" is a picture of a microprocessor showing less detail  than ``cpu B". ``Escher" is a painting of tiles by Escher. ``Paper" is a corrugated sheet of paper. ``Shadows B" are the shadows produced by sea tides. ``Symmetric B" is ``shadows B" repeated 4 times. ``Symmetric A" is ``symmetric B" repeated 4 times. ``Rectangle C" is ``rectangle B" repeated 4 times. Those images tagged  ``Peano curve" are space filling curves. ``Periodic" and ``Alternated" are of a similar type, and we thought they would have low complexity, being simple images. ``Random" was a computer-generated image, technically pseudo-random. This randomly-generated picture will serve to illustrate the main known difference between algorithmic complexity and logical depth. i.e., based on its algorithmic (program-size) complexity the random image should be ranked at the top, whereas its logical depth would place it at the bottom. All pictures were randomly chosen from the web, transformed to B\&W and reduced to 600 $\times$ 600 pixels\footnote{Unlike the printed version, if seen in electronic form the images can be zoomed in, allowing better visualization. They are also available online under the paper title at \url{http://www.algorithmicnature.org}}.

\subsection{Towards the choice of lossless compression algorithm}

Deflate\footnote{RFC1951: Deflate Compressed Data Format Specification version 1.3 \url{http://www.w3.org/Graphics/PNG/RFC-1951}.} is a compliant lossless compressor and decompressor available within the zlib package. Deflate compression compresses data using a combination of the Lempel-Ziv coding algorithm \cite{ziv} and  Huffman coding \cite{huffman}. It is one of the most widely-used compression encoding systems.

 Huffman coding assigns short codewords to those input blocks with high probabilities and long codewords to those with low probabilities. In other words, the compressor encodes more frequent sequences with a few bytes and spends more bytes only on rare sequences. 

The Lempel-Ziv coding algorithm builds a dictionary and encodes the string by blocks using symbols in the dictionary. The Lempel-Ziv algorithm leads to actual compression when the input data is sufficiently large and is characterized by sufficient redundancy (patterns)  \cite{welch}. The Lempel-Ziv compression algorithm provides upper bounds to $K$.

\subsection{Compression method}
\label{comp}

Since digital images are just strings of values arranged in a special way, one can use image compression techniques to approximate the algorithmic complexity of an image. And through its image, the algorithmic complexity of the object represented by the image (at the scale and level of detail of the said image). 

A representative sample of lossless image-compression algorithms (GIF, TIFF and JPG 2000) was tested in order to compare the decompression runtimes of the images being studied. Although we experimented with these lossless compression algorithms, we ended up choosing PNG for several reasons, including its stability and the flexibility it affords, given that it permits the use of open-source developed optimizers for further compression.

 Portable Network Graphics (PNG) is a bitmapped image format that employs lossless data compression. It uses a 2-stage compression process, a pre-compression or filtering, and Deflate. The 
filtering process is particularly important in relating separate rows, since Deflate on its own has no understanding that an image is a bi-dimensional array, and instead just sees the image data as a stream of bytes.

 There are several types of filters embedded in image compression algorithms which exploit regularities found in a stream of data based on the corresponding byte of the pixel to the left, above, above and to the left, or any combination thereof, and encode the difference between the predicted value and the actual value. The number of combinations tested is finite and limited by the compression algorithm. By applying different filters a lossless image compression algorithm uses the data contained in the pixels $x_1, \ldots, x_8$ to predict the value of another pixel $x_9$ (see Figure 3) that may save space, enabling the compression of the said image without losing any information. Yet one can optimize the search for a successful combination by setting the compression algorithm in such a way that it tries harder, and spend more time trying to more effectively compress the data.

\begin{figure}[h!]
  \centering
      \scalebox{.4}{\includegraphics{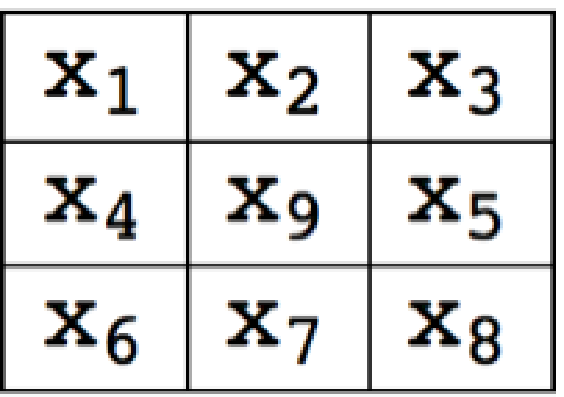}}
  \caption{Image pixel neighborhood. }
\end{figure}

Compression can be further improved by so-called PNG-optimizers using 
more filter methods and several other lossless data compression
 algorithms. Among these optimizers 
are Pngcrush\footnote{Syntax example: pngcrush -reduce -brute -e ``.compressed.png" /testimages/*.png. More info: \url{http://pmt.sourceforge.net/pngcrush/}} and
AdvanceCOMP\footnote{Syntax example: e.g. advdef -z -4 *.png ('4' indicating the 
so-called ``insane" compression according to the developers). More info: \url{http://advancemame.sourceforge.net/}}, two of the most popular open-source optimizers. They tried various compression methods and  were able to reduce the PNG files by about 10 to 20\% of their original length. AdvanceCOMP recompresses PNG files (and other file formats) using the Deflate 7-Zip implementation. The 7-Zip Deflate encoder effectively extends the
 scope of Deflate further by performing a much more detailed search of
compression possibilities at the expense of significant further
 processor time spent searching, which for this experiment was not a 
matter of concern. 7-Zip Deflate also uses the LZMA algorithm, an improved and 
optimized version of the LZ77 \cite{ziv} compression algorithm. The LZMA\footnote{More technical details are given in \url{http://www.7-zip.org/7z.html}.} algorithm divides the data into 
packets, each packet describing either a single byte or an LZ77
sequence, with its length and distance implicitly or explicitly
 encoded\footnote{A useful website showing a benchmark of compression 
algorithms is at url{http://tukaani.org/lzma/benchmarks.html}.}.

 As a sort of  verification, we ran a popular zip-based commercial compressor, set to the maximum possible compression, over the already compressed and optimized files. The zip-based archiver was unable to further compress any of the files (they were actually always a little larger in size).

\subsection{Using decompression times to estimate complexity}

Inflate is the decoding process that takes a Deflate bit stream for decompression and correctly produces the original full-size data file. To decode an LZW-compressed file, one needs to know the dictionary encoding the corresponding strings in the original data to be reconstructed. The decoding algorithm for LZ77 works by reading a value from the encoded input and outputting the corresponding string from the shorter file description.

 It is this decoding information that Inflate uses when importing a PNG image for display, so the lengthier the directions for decoding, the longer the time it takes. We are interested in these compression/decompression processes, particularly the compression size and the decompression time, as an approximation of the algorithmic complexity and the logical depth of an image.

 The decompression directions for trivial or random-looking objects are simple to follow, with the decompression process taking only a small amount of time, simply because the compression algorithm either compresses very well and the decompression is just straightforward (a trivial case), or does not compress at all (a random case). Longer runtimes, however, are usually the result of a process following a set of time-consuming decompression instructions, hence a complex process.

\subsection{Timing method}

The execution time was given by the \emph{Mathematica} function Timing\footnote{Timed on two different computers for validation: on a MacBook Intel Core 2 Duo 2GHz, 2048MB DDR2 667Mhz with a solid-state
 drive (SSD) and on a MacBook Pro Intel Core 2 Duo 2.26Ghz, 4096MB DDR3 1067Mhz with a traditional hard disk drive (HDD), both running Mac OS X Version 10.6.1 (Snow Leopard). The MacBook Pro was always twice as fast on average.}. The function Timing evaluates an expression and returns a list of the time taken in seconds, together with the result obtained. The function includes only CPU time spent in the \emph{Mathematica} kernel.

The fact that several processes run concurrently in computing systems as part of their normal operation is one of the greatest difficulties faced in attempting to  measure with accuracy the time that a 
decompression process takes, even when it is isolated from all other computer processes. This 
instability is due to the fact that the internal operation of the computer comes in several layers, mostly at the operating system level. In order to avoid measurement perturbations as much as possible, several stabilizing measures were undertaken:

\begin{myitemize}
\item Most computer batch processes and operating system services were disabled, including services such as wireless, bluetooth and energy saving features, such as the hard drive sleeping and display dimming features\footnote{To understand the number of layers of complexity involved in a modern computing system see Tanenbaum's book on operating systems \cite{tanenbaum}.}.
\item The microprocessor was warmed up before each experiment by running an equivalent process (e.g. a mock experiment run) before running the actual one, in order to have the fan and everything else 
already working optimally (like preheating an oven).
\item The cache memory was cleared after each function call using the \emph{Mathematica} function ClearSystemCache. No history was saved in RAM memory.
\item A different order of images was used for each experiment run. The result was averaged with at least 30 runs, each compressing the images in a different random order. This helped to define a confidence level on the basis of the standard deviation of the runs-- when normally distributed-- a confidence level with which we were satisfied and at which we arrived in a reasonable amount of time. Further runs showed no further improvement. Measurement stabilization was reached after about 20 to 30 runs.
\end{myitemize}

The presence of some perturbations in the time measure values was unavoidable due to lack of complete control over all the computing system parameters. As the following battery of tests will show, one can reduce and statistically curtail the impact of this uncertainty to reasonable levels.

%\pagebreak

\section{Results}
\label{results}

\subsection{Controlled experiments}

\subsubsection{Image size uncertainty variation}
\label{size}

\begin{figure}[h!]
  \centering
      \scalebox{.62}{\includegraphics{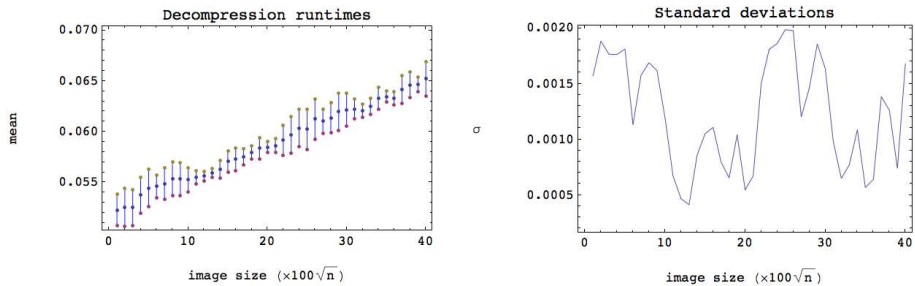}}
  \caption{Decompression times and standard deviations for monochromatic images increasing in size.}
\end{figure}

\begin{figure}[h!]
  \centering
      \scalebox{.62}{\includegraphics{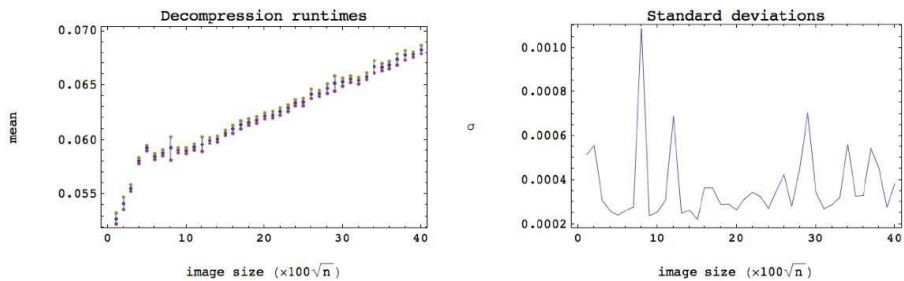}}
  \caption{Decompression times and standard deviations for images with (pseudo) random noise increasing in size.}
\end{figure}

Figures 4 and 5 show that decompression times using the PNG algorithm and the optimizers increase in linear fashion when image sizes increase linearly. The distributions of standard deviations in the same figures show no particular tendency apart from suggesting that the standard deviations are likely due to random perturbations and not to a bias in the methodology. 

A comparison with Figure 6 shows that standard deviations for images containing random noise always remained low, while for uniformly colored images standard deviations were significantly higher.

\begin{figure}[h!]
  \centering
      \scalebox{.7}{\includegraphics{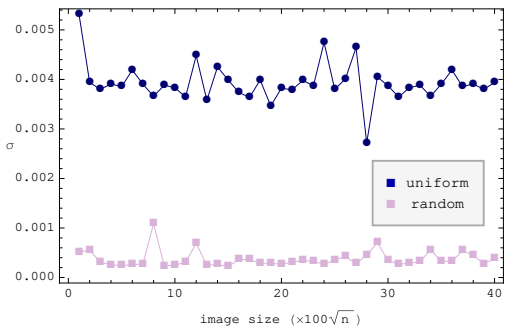}}
  \caption{For uniformly colored images standard deviations of decompression times were larger on average. For random images the standard deviations were smaller and compact. The size of the image seemed to have no impact on the behavior of the standard deviations in either case.}
\end{figure}

\begin{figure}[h!]
  \centering
      \scalebox{.7}{\includegraphics{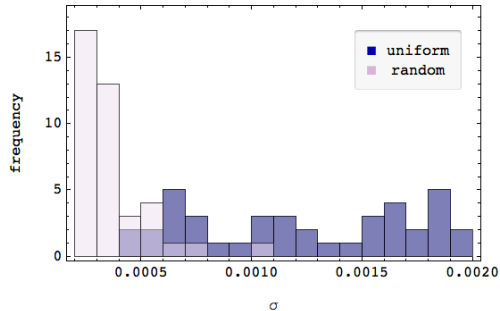}}
  \caption{Histogram of the standard deviations for uniformly colored and random images increasing in size.}
\end{figure}

\subsubsection{Fully controlled test}
\label{cuadros}

For this test only, we used a toy compression algorithm designed to be as simple as possible in order to estimate the uncertainty due to system perturbations. We devised a test in which we would have full control of the data and the compression algorithm. By understanding the processes we expected to be better able to predict compression rates and decompression uncertainties, and to quantify the uncertainty of the measurement of decompression times in the general case (i.e. using other compression algorithms). 

It was found that the algorithmic complexity approximation (the file size) decreased linearly at a rate of 1990 bits per image, the same as the number of regular inserted bits per image, taking about $\log_2(2000)$ to encode the compressed regular string, as would have been theoretically predicted (by Shannon's information theory). One can also predict the decompression runtime by calculating the slope of the decompression rate. Each insertion of 2000 regular bits into the random image took 0.00356 seconds less than the preceding one, fitting the estimate computed by the rate ratio. 

The behavior of the standard deviations suggests that the more random an image the less stable the decompression time, and the more regular the more stable. This may be explained by the number of operations that the toy compression algorithm uses for encoding a regular string. In the extreme case of a uniformly colored image, the code comes up with a single loop operation to reproduce the image, taking only one unit of time. On the other hand, when an image is random, wholly or partially, the number of 
operations is larger because the algorithm finds small patterns everywhere, patterns that have a timing cost per operation performed. Each loop operation using the Table function in \emph{Mathematica} takes
 0.000025 seconds on average, while each iteration takes only about 0.00012 seconds on average, suggesting that it is the number of operations that determines the runtime.

\begin{figure}[h!]
  \centering
      \scalebox{.56} {\includegraphics{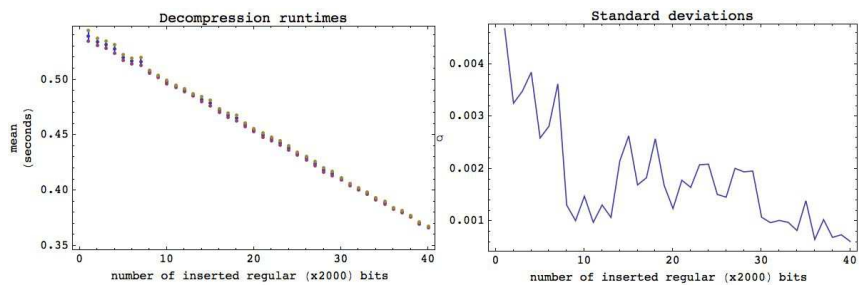}}
  \caption{Using the toy compression algorithm, the larger the white region in an image with random noise, the shorter the decompression runtime and the lower the standard deviations.}
\end{figure} 

The standard deviation of the mean 0.0000728 is smaller than the average of the standard deviations of each point, which is 0.000381. This shows that runtime perturbations were not significant enough.

\subsubsection{Using the PNG compression algorithm}

This time we proceeded to apply the PNG compression algorithm together with the optimizers (Pngcrush and AdvancedCOMP) to the same computer-generated images with random noise used in the previous section, confident that we understood the variables involved in the previous experiment, and that everything seemed to be controlled.

\begin{figure}[h!]
  \centering
      \scalebox{.6}{\includegraphics{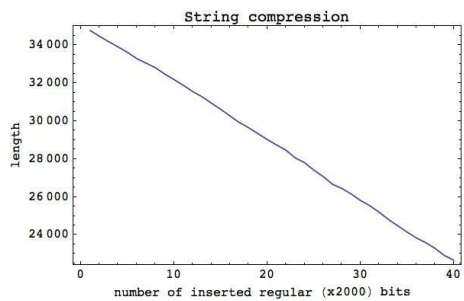}}
  \caption{Compression rate for uniformly colored images. As for the controlled experiment with the toy compression algorithm using the PNG compression algorithm, the compressed length of the images also decreased in a predictable way, as one would have expected for such a controlled experiment. And they did so at a rate of 310 bits per image.}
\end{figure}

Figure 9 shows that the PNG\footnote{with Pngcrush and AdvanceCOMP} compression algorithm followed the same trend as the toy compression algorithm, namely the more uniformity in a random image the greater the compression. PNG however achieved greater compression ratios compared to the toy compression algorithm, for obvious reasons.

%\begin{figure}[h!]
 % \centering
    %  \scalebox{.5}{\includegraphics{anomaliesplot.eps}}
 % \caption{Magnified anomalies found in the decompression times deviating from the main slope, given by $\Delta(y) = y - (b + a (x - 1))$, seem to behave randomly, suggesting no methodological bias.}
%\end{figure}

In Figure 9 one can see that there are some minor bumps and hollows all along. Their deviation does not, however, seem significantly different from a uniform distribution, as shown in Figure 10, suggesting bias toward no particular direction. As was the case with the toy compression algorithm, using the PNG algorithm too caused algorithmic complexity to decrease as expected--upon the insertion of uniform strings.

\begin{figure}[h!]
  \centering
      \scalebox{.56}{\includegraphics{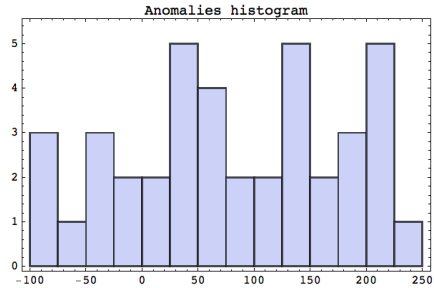}}
  \caption{Anomalies found in the decompression times deviating from the main slope  seem to behave randomly, suggesting no methodological bias.}
\end{figure}

\begin{figure}[h!]
  \centering
      \scalebox{.52}{\includegraphics{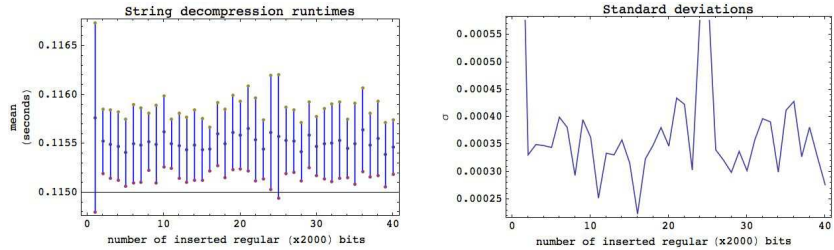}}
  \caption{Random images produce smaller standard deviations (notice the plot scale). But unlike the toy case, decompression times remained statistically the same despite an increase in image size.}
\end{figure}

\subsubsection{Increasing complexity test}
\label{lineas}

\begin{figure}[h!]
  \centering
      \scalebox{.64}{\includegraphics{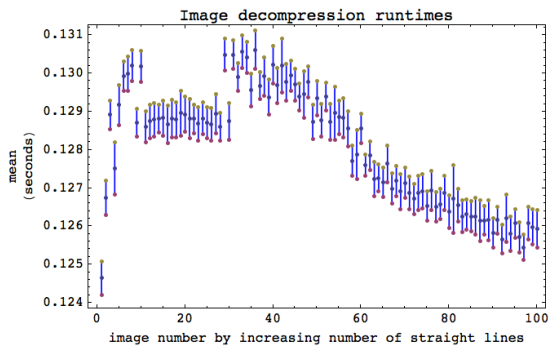}}
  \caption{Compressed path: Going from all-white to all-black by randomly inserting black lines.}
\end{figure}

The first thing worth noticing in Figure 11 (left) is that the standard deviations remained the same on average, suggesting that they were not related to the image, but rather to external processes. The process of randomly depicting lines seems to have a low maximum limit of complexity, which is rapidly reached. Nevertheless, eight bins of images with significantly different decompression time values, that is, with non-overlapping standard deviations, were identified. What Figure 12 suggests is that increasing the number of lines can only lead to a limited maximum degree of complexity bounded by a convex curve. The eight significantly different images have been selected according to jumps that were maximizing their differences: $n = \Delta D / \sigma$ with $\Delta=\max\{D(I_i) | i \in \{1, \ldots, n\}\} - \min\{D(I_i) | i \in \{1, \ldots, n\}\}$, with $I_i$ each image in the set, and $D$ the logical depth. 

As regards the discontinuity in the graph, one hypothesis is that the compression algorithm reaches a threshold favoring some regularities over others, producing jumps in the decompression times. For a certain quantity of randomly depicted lines, the limit remains stable for a while once reached, until the moment when the increasing number of inserted lines fills up the space and the configuration reaches its lowest complexity by decompression time, at which point it begins to approach a phase in which it resembles a uniformly colored image.

 The decompression time depicted in Figure 12 turned out to be very interesting, suggesting what one might expect for this kind of experiment: a path traced between two monochromatic (fully colored) images, an initial all-white image succeeded by images of increasing complexity comprising random lines, and ending at the horizontal departure line in a final monochromatic, almost all-black image, when the space becomes entirely filled with black lines.

\subsection{Calibration tests}
\label{calibration}

The following are the classifications of series 1, 2 and 3 according to their decompression runtimes as described in the methodology section \ref{method}.

\begin{figure}[h!]
  \centering
   \scalebox{0.56}{\includegraphics{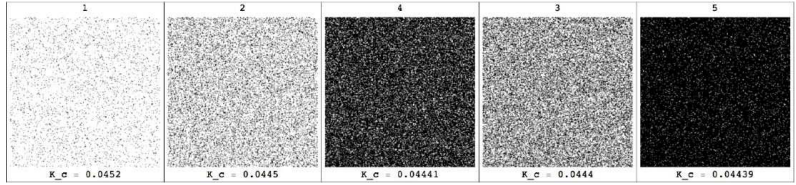}}
  \caption{Series 1 classification: Random points with different densities.}
\end{figure}

% [width=0.5\textwidth]

\begin{figure}[h!]
  \centering
   \scalebox{0.48}{\includegraphics{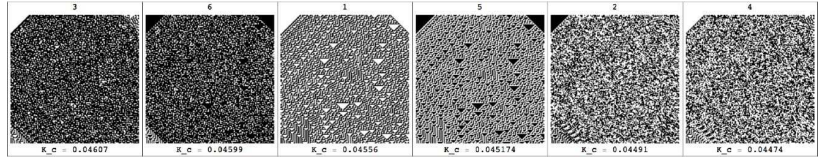}}
  \caption{Series 2 classification: Cellular automata superpositions, rotations and inversions.}
\end{figure}

\begin{figure}[h!]
  \centering
   \scalebox{0.5}{\includegraphics{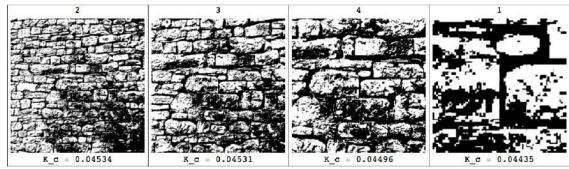}}
  \caption{Series 3 classification: Zooming in a wall.}
\end{figure}

\begin{myitemize}
\item Series 1 classification (Figure 13): Images labeled 1 and 2 never came last, while images 4 and 5 never came first. The first half of the images remained the same run after run, while the second half also remained stable, with some occasional permutations among the intermediate elements of each half, but never with the extremes exchanging places.
\item Series 2 classification (Figure 14): A cellular automaton superposed upon itself rotated 90${}^{\circ}$, with random bits inserted. The runs not only mostly never changed, the procedure consistently sorted the images into pairs, grouping images with similar characteristics (to which the same function was applied).
 The classification seems consonant with what one may take to be the more random vs. the less random. All of them had low physical complexity in general, given their values, which accords with what is believed about cellular automaton rule 30, i.e. that it is known to behave randomly \cite{wolfram}, with 
some regular structures on the left.
\item Series 3 classification (Figure 15): A wall and the same wall but viewed from a closer vantage point. It turned out to be highly stable after several runs, sorting the series of images from the farthest to the closest picture of the wall, suggesting that pictures taken from a greater distance captured more of the structure of the wall, while zooming in without also increasing the resolution meant losing detail and structure. The closest picture always came last, while the farthest always came first. There were only occasional permutations among those in between.
\end{myitemize}

\begin{table}
\label{deviations}
\begin{center}
$
\begin{array}{|c|c|c|}
\hline
\text{series number}&\text{description}& \textit{Std. deviation} \\
\hline
2 & \text{cellular automata} & 0.000556 \\
3 & \text{wall zooming} & 0.00046 \\
1 & \text{random points} & 0.000349 \\
\hline
\end{array}
$
\end{center}
  \caption{Standard deviations per series sorted from greatest to lowest standard deviation.}
\end{table}

As was expected (see section \ref{method}), a greater standard deviation for the series of cellular automata was found due to the image pairing. Each image occurred in tandem with its inverse, but permutations were common--and expected--between members of pairs, which intuition tells us ought to have the same complexity. The expected permutations produced a larger standard deviation.

\subsection{Compression length ranking}
\label{main}

The following experiments were carried out as described in section \ref{method} using 56 black-and-white images spanning a range of objects, each seemingly having a different complexity which we could intuitively gauge more or less accurately\footnote{The images are available online under the paper title at \url{http://www.algorithmicnature.org}}. Each image has a very short description, followed by the image itself and by the approximate values of $K_c$ (Figure 16) and $D_c$ (Figure 17) for our general compression algorithm $c$.

\begin{figure}[h!]
  \centering
      \scalebox{.65}{\includegraphics{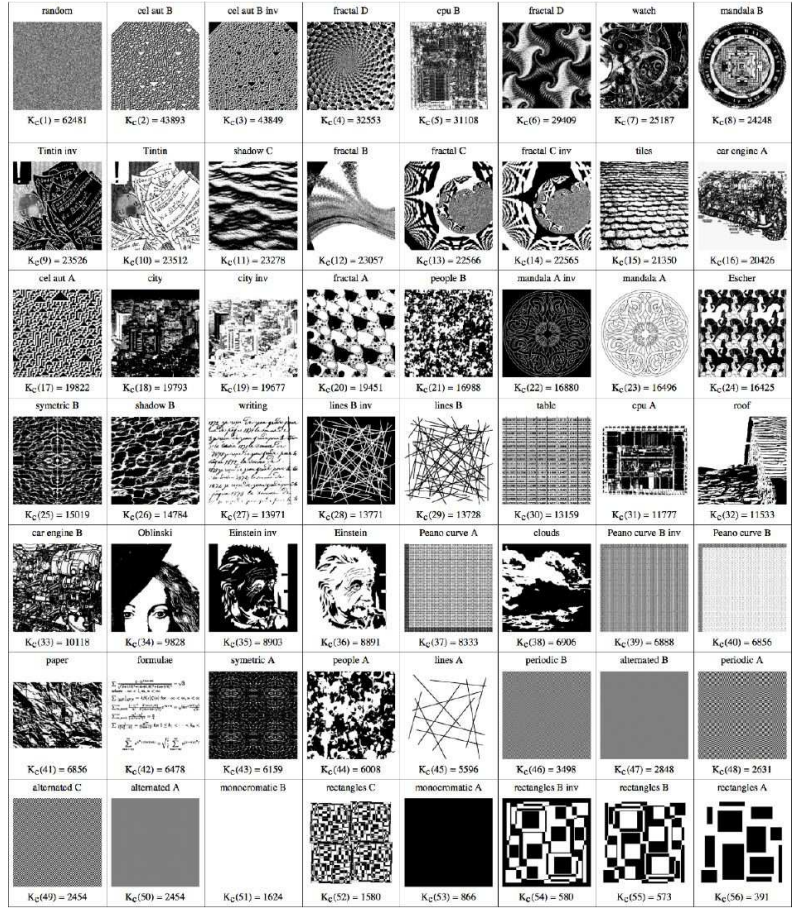}}
  \caption{Ranking by compressed lengths: $K_c(i_n) \leq K_c(i_{n+1})$ with $K_c(i)$ the approximation to the algorithmic complexity $K(i)$ of the image $i$ by means of the PNG compression algorithm aided by Pngcrush and AdvanceCOMP. These classifications are available online for better visualization at \url{http://www.algorithmicnature.org} under the paper title.}
\end{figure}

The classification in Figure 16 presents the images ranked according to their compressed lengths using PNG image compression together with the PNG optimizers. It goes from larger to smaller, and as can be seen, the more random-looking or highly structured, the better classified, while trivial images come last. One can verify that the procedure is invariant to simple transformations, such as inversions and complementations, since images and their symmetric versions are always next to (or close to) each other, indicating that the compression algorithm behaves as one would expect (that is, that inverting colors, for example, has no impact on the resultant compressed length). One would hardly say, however, that 
a random image is physically complex (random processes, like a gas filling a room, unfold in almost zero time and seem to require no great computational power), which is why plain algorithmic complexity does not help to distinguish between complexity associated with randomness and complexity associated with highly structured objects.

\subsection{Decompression times ranking}

\begin{figure}[h!]
  \centering
      \scalebox{.65}{\includegraphics{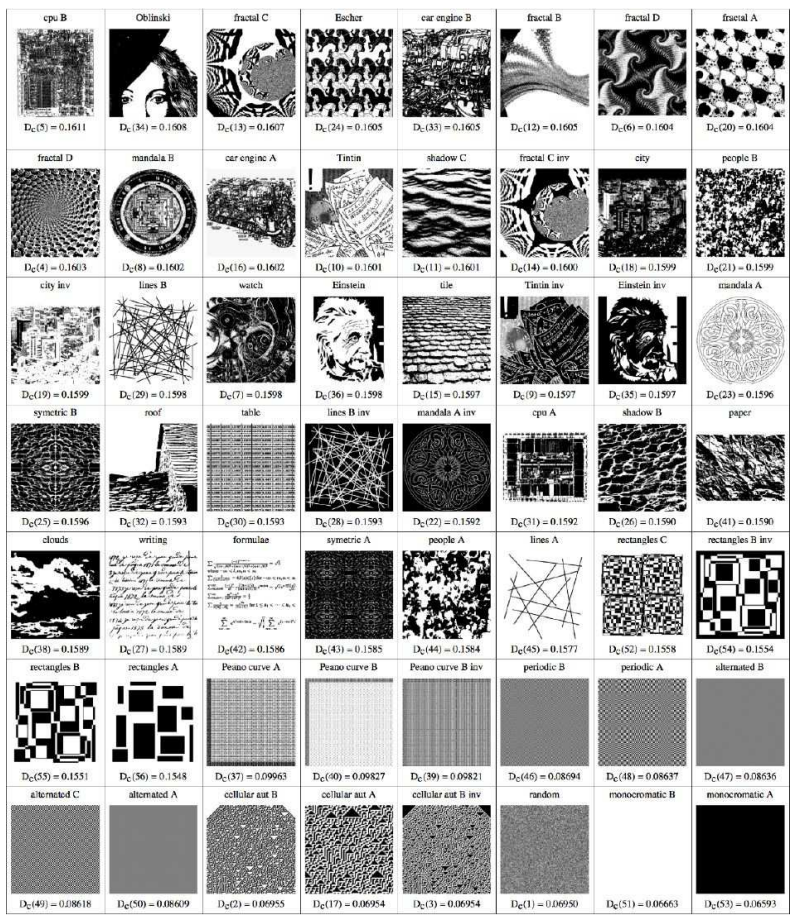}}
  \caption{Ranking by decompression times: $D_c(i_n) \leq D_c(i_{n+1})$ with $D_c(i)$ the approximation to the logical depth $D(i)$ of the image with number $i$ according to the indexing from the previous classification for $K_c$, by means of the PNG compression algorithm aided by Pngcrush and AdvanceCOMP.}
\end{figure}

The classification in Figure 17 goes from greater to smaller logical depth based on the decompression runtimes. The number at the bottom is the time in seconds that the PNG algorithm took to decompress the images when applied to their compressed versions.
 Images we gauged to have the highest physical complexity came first. It is also worth mentioning that the images and their inverted versions remained close to each other, meaning that they were always in the same complexity group, which too is just what we expected. This also indicates the soundness of the procedure, since images and their inversions should be equal in complexity, and therefore equal in complexity to the object as well, at a commensurate scale and level of detail.

\begin{figure}[h!]
  \centering
      \scalebox{.62}{\includegraphics{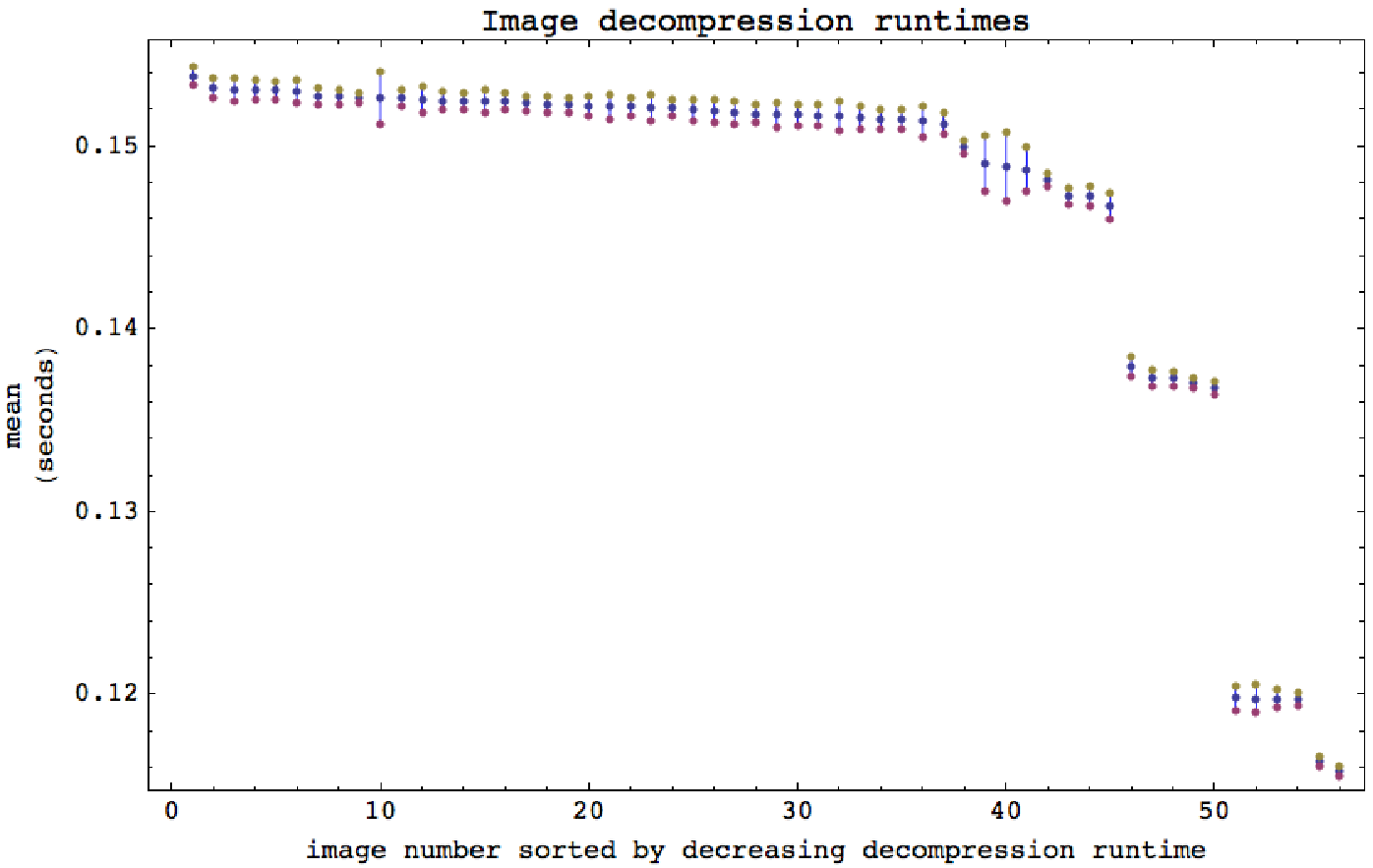}}
  \caption{Mean and standard deviations of the decompression times of
 the 56 images.}
\end{figure}

Figure 18 shows some of the jumps seen before in the experiments in section \ref{lineas}. We think they may be due to the behavior of the compression algorithm. The compression algorithm applies several filters, and it may favor some regularities over others that are better (faster) decoded after a certain threshold is reached, producing these jumps.

\begin{figure}[h!]
  \centering
      \scalebox{.75}{\includegraphics{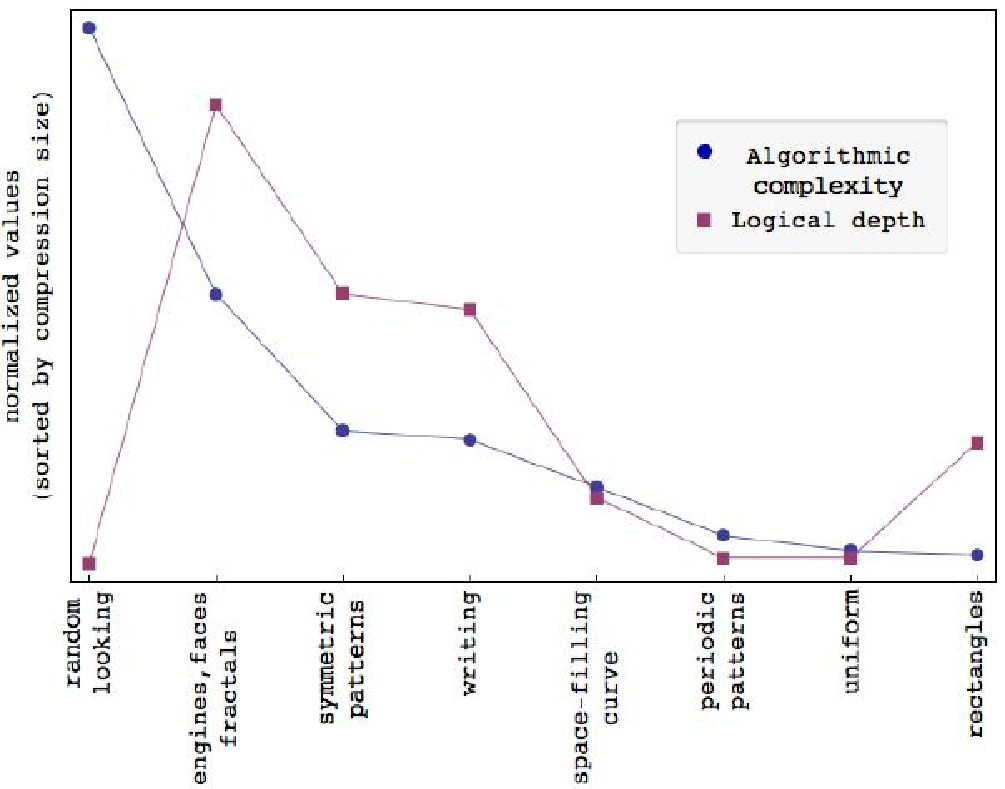}}
  \caption{Rank comparison between $K_c$ and $D_c$ as functions of the chosen images 
having significantly different $D_c$ values.}
\end{figure}

\begin{figure}[h!]
  \centering
      \scalebox{.7}{\includegraphics{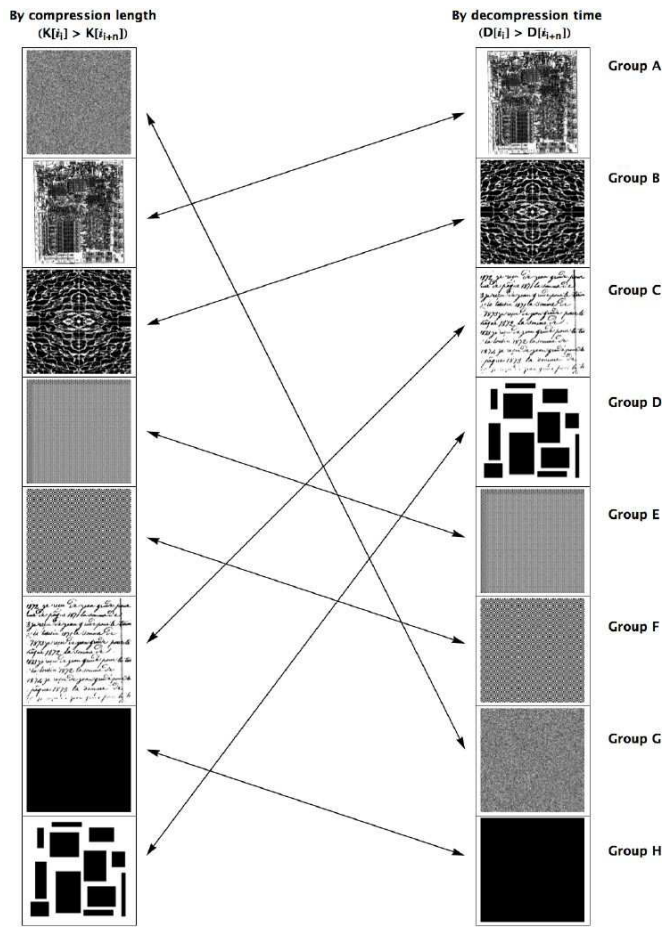}}
  \caption{$K$ to $D$ mapping by groups indicating the complexity group (by logical depth) to which each image instance on the right belongs.}
\end{figure}

\begin{figure}[h!]
  \centering
      \scalebox{.9}{\includegraphics{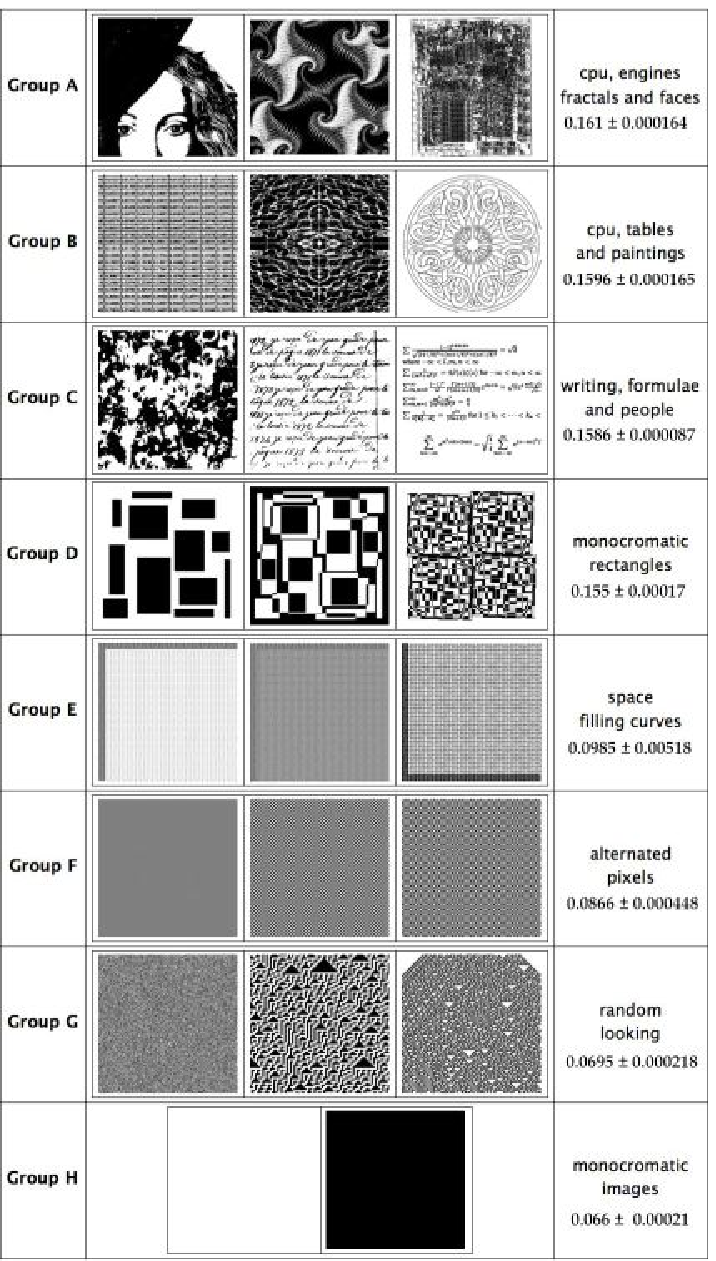}}
  \caption{Significantly different groups by decreasing decompression time.}
\end{figure}

Images were grouped in 8 significantly different groups (with different decompression times and therefore seemingly different logical depths). Formally, $\bar{x}(g_i) \pm \sigma(g_i) > \bar{x}(g_j) \pm \sigma(g_j)$ for any two different group indexes $i, j \in \{A, B, \ldots, H\}$.

 The average difference between the largest and the smallest $K_c$ value was about $62090$ bits, while the average difference between the largest and the smallest $D_c$ was $0.095$ seconds. The largest calculated compressed image size (image 1) was 159.8 times larger than the shortest compressed image size (image 56). The largest evaluated decompression time (image 5) was 2.46 times greater than the shortest calculated decompression time (image 53). 

No significant statistical correlation between $K_c$ and $D_c$ was found, indicating that  $K_c$ and $D_c$ are actually two different measures. Figures 19 and 20 illustrate the classification values from $K_c$ to 8 significantly different decompression time ($D_c$) groups according to each of the two measures. The ranking of images based on their decompression times corresponds to the intuitive ranking resulting from a visual inspection, with things like microprocessors, human faces, cities, engines and fractals figuring at the top as the most complex objects, as shown in Figure 21 (group A), while random-looking images, ranked high by algorithmic complexity, were ranked low (group G) according to the logical depth expectation, being classified next to trivial images such as the uniformly colored (group H), and thus indicating the characteristic feature of the measure of logical depth. A gradation of different complexities can be found in the groups between, gradually increasing in complexity from bottom to top.

\section{Conclusions and further work}
\label{conclusions}

Extensive experiments were conducted. Along the way we think we have shown that:

\begin{myenumerate}
\item Ideas in the spirit of Bennett's logical depth can be implemented to approach a real-world characterization and classification problem, instead of simply using the concept of algorithmic complexity alone, which distinguishes only between random and trivial objects but does not separate random from structured objects.
\item Based on the study of several cases and the testing of several compression algorithms, the method described in this paper has been shown to work, and to be of use in identifying and classifying images by their apparent physical complexity.
\item The procedure described herein constitutes an unsupervised method for evaluating the information content of an image by physical complexity.
\item The method involving the use of decompression times yields a reasonable measure of complexity that is different from the measure obtained by considering algorithmic complexity alone, while being in accordance with intuitive expectations of greater and lesser complexity.
\end{myenumerate}

 It remains to investigate how this procedure may be improved upon to accurately follow Bennett's stricter definitions of Logical Depth. For example, it could be that by allowing a couple of additional bits in the compressed versions of the images one gets shorter decompression times, and we think experiments in this connection are possible. To stay true to the spirit of Bennett, one may use several compressors $c_1, \ldots, c_n$ on strings $s_1, \ldots, s_m$ and compute $D_{c_i(s_j)}$ for all $1 \leq i \leq n$ and $1 \leq j \leq m$.

 Then, an object $s_h$ is assumed to be more complex than an object $s_k$ if $D_{c_i(x_j)} > D_{c_i(x_k)}$ for all $1 \leq i \leq n$ and a partial order is thus obtained, from which a total order could be obtained by calculating the (e.g. harmonic) average of the decompression times.

\section*{Acknowledgments}

The authors would like to acknowledge the anonymous referees whose generous comments have helped us improve this paper.

\end{document}